\documentclass[final,3p,times,twocolumn,numbers]{elsarticle}

\biboptions{square,numbers}

\usepackage{amssymb}
\usepackage{amsmath}
\usepackage{lipsum}
\usepackage{booktabs}
\biboptions{sort&compress}

\journal{Physics Letters B}

\begin{document}

\begin{frontmatter}



\title{A Note on a High-Energy Limit of Gravitational Perturbation Theory}


\author[aff]{Poul H. Damgaard}
\author[aff]{and Ludovic Plant{\'e}}
\affiliation[aff]{organization={Niels Bohr International Academy, Niels Bohr Institute, University of Copenhagen},
            addressline={Blegdamsvej 17, Copenhagen DK-2100}, 
            country=Denmark}

\begin{abstract}
The post-Minkowskian expansion of general relativity expands the metric in Newton's constant G around flat space-time, but much effort currently focuses on ways to extend its range of applicability by summing the perturbative series to all orders. A regime of particular interest is a specific high-energy limit where a formula derived recently provides the momentum kick of binary scattering problem to all orders, and even suggests an approach towards strong coupling. We show that this is an asymptotic expansion. Although the series diverges we give evidence that it is Borel summable, and the full function in integral form is finite. As a by-product we determine the transcendentality properties of the coefficients in the expanded expression.

\end{abstract}







\end{frontmatter}




\section{Introduction}
\label{introduction}

Is the post-Minkowskian expansion in general relativity convergent or divergent? It might be tempting to extrapolate from more than a century of experience with perturbative expansions in quantum mechanics and quantum field theory to conclude that weak-coupling perturbation theory in Newton's constant $G$ is asymptotic only.
However, the reasons for perturbation series divergences in the quantum field theoretical framework do not carry over to the classical theory of gravity. Indeed, it is possible to solve the classical Einstein equations of motion perturbatively for the Schwarzschild metric in harmonic gauge. Classical recursion relations of this post-Minkowskian expansion can push the calculation to any order, and it is possible to prove by induction the form of the expansion to all orders in $G$. The sum is a geometric series, convergent for all radii $r > GM$ as expected, and its sum equals the standard harmonic-gauge expression for the Schwarzschild metric in that region \cite{Damgaard:2024fqj}. A similar perturbative expansion of the Kerr metric can be performed based on the same classical recursion relations, and it can also be pushed to arbitrarily high order, agreeing at every step with the standard non-perturbative expression expanded \cite{Damgaard:2026ocb}. There are thus explicit counterexamples to the folklore that the post-Minkowskian expansions are always asymptotic at best. One might even hope that the usual weak-coupling expansion of the gravitational two-body problem, at least in suitably defined subsectors, may be convergent as well \cite{Damgaard:2026whk}.

Such hopes of perturbative resummations underpin several new directions in the field \cite{Cheung:2023lnj,Cheung:2024byb,Rothstein:2024nlq,Guevara:2024edh,Akpinar:2025huz,Alessio:2025isu,Kosmopoulos:2025rfj,Alessio:2026bdi,Barack:2026izc,Chang:2026eti,Solon:2026ubm}. A major step forward has very recently been taken by Saavedra \cite{Saavedra:2026ttk} who proposes (and checks, by comparing with existing fixed-order calculations in $G$) a regime at high energy that permits a precise resummation {\em to all orders in $G$} of a subset of diagrams contributing to the gravitational momentum kick in the two-body problem. This is remarkable in at least two ways: (1) any consistent (read: exact) resummation of the perturbative expansion to all orders is a leap forward in itself, and (2) the regime in which this resummation is argued to be exact edges very close to that of strong coupling, despite all ingredients ab initio being derived from weak-coupling perturbation theory. This will be briefly reviewed below. It resembles the phenomenon of exact resummation of the Schwarzschild metric all the way up to the harmonic-gauge coordinate singularity, as was shown in ref. \cite{Damgaard:2024fqj}.

While a full analytical resummation of the post-Minkowskian expansion for the gravitational two-body problem may seem completely out of reach, one may still search for kinematical regimes where gauge-invariant subsets of the expansion can be cleanly separated out. This strategy is precisely what is behind the new work of ref. \cite{Saavedra:2026ttk}. Such a developments reaches back to, and connect to, seminal work in the theory of the gravitational eikonal, and in particular to refs. \cite{Amati:1990xe,Amati:1993tb,Gruzinov:2014moa,Ciafaloni:2015xsr}. Indeed, by careful tunings of limits the surprisingly compact formula for high-energy gravitational radiation loss of Gruzinov and Veneziano \cite{Gruzinov:2014moa} emerges from ref. \cite{Saavedra:2026ttk}, thus providing a first highly non-trivial check.

The purpose of the present paper is to analyze the convergence properties of the series expansions derived in ref. \cite{Saavedra:2026ttk}. We should from the outset be clear about the implications of this: the infinite-order resummation is not in $G$ but in rescaled couplings $\alpha_i$ (one for each participating scatterer) that are proportional to $G$ but which also depend on energy, mass, and impact parameter. So one cannot draw more general conclusions from this, but we find it important to settle nevertheless. To jump to our conclusion, we find that the all-order expansion of ref. \cite{Saavedra:2026ttk} can be given by analytic expressions to arbitrarily high order, and that it diverges. It is an asymptotic series, and we shall give evidence that it is Borel summable. The un-expanded closed-form expression is finite, and it agrees with the Borel-summed result. In addition we sharpen a conjecture of \cite{Saavedra:2026ttk} regarding the transcendentality of the coefficients of the expansion.

\section{The high-energy limit}
Here we briefly review the set-up and results of ref. \cite{Saavedra:2026ttk}, using the same notation as that paper (with only one exception, see below). Consider the gravitational center of mass scattering of particle 1 of mass $m_1$ and incoming momentum $p_1$ against particle 2 of mass $m_2$ and incoming momentum $p_2$. Because of the high-energy limit it is convenient to use light-cone coordinates $p_1^{\mu} = (p_1^+,p_1^-, q^{\mu}/2)$ and $p_1^{\mu} = (p_2^+,p_2^-, -q^{\mu}/2)$, with $q^{\mu} = p_1^{\mu} - p_2^{\mu}$ being the momentum transfer. We define $\gamma = p_1\cdot p_2/(m_1m_2)$, which is the relative $\gamma$-factor in the center of mass frame (here we differ in notation from ref. \cite{Saavedra:2026ttk}). 

The high-energy limit considered in \cite{Saavedra:2026ttk} is based on defining $\alpha_i = Gs/(m_ib)$ where $s = (p_1 + p_2)^2 = p_1^-p_2^+$ and $b$ is the magnitude of the impact parameter. In the usual post-Minkowskian expansion the effective dimensionless coupling would be $Gm/b$ or $G\sqrt{s}/b$ in the high-energy or massless limits. Ref. \cite{Saavedra:2026ttk} chooses as the effective expansion parameter $m/\sqrt{s}$ at finite $\alpha_i$. Remarkably, in this sweet spot a scaling regime exists where $\alpha_i$ is kept fixed, $s$ is large, and the expansion is arranged in powers of $m^2/s$. Since the analysis is perturbative, $\alpha_i$ is limited by $Gs/(mb) \sim 1$ but this is still tantalizingly inching towards strong coupling in the high energy scattering, and we have more to say below.

We are now ready to import the main results of Saavedra's analysis. We restrict ourselves to the minus-component of the momentum kick of, say, particle 1, $i.e$ $\Delta p_1^-$; the other components can be treated analogously, as explained in \cite{Saavedra:2026ttk}. Correcting what appears to be a misprint in eq. (11) of that paper, we use as an intermediate step that the leading piece of the radiation kernel becomes
\begin{equation}
{\mathcal N}(b',y) = \frac{G m_1^4}{\pi^2}\,y^{2}\,K_2(-ym_1|b'|)^2
\end{equation} 
where $K_2(x)$ is a modified Bessel function, $y = \ell^-/p_1^-$ for momentum $\ell^{\mu}$ of the graviton, and $b'$ is the transverse "impact parameter" of the graviton, the Fourier dual of the transverse part of $\ell^{\mu}$. This leads to a very compact integral representation for $\Delta p_1^-$,
\begin{eqnarray}
 && \frac{p_2^+}{m_1^2}\,\Delta p_1^- =
-4\alpha_1^{2} \cr
&& - \frac{4\, s}{m_1^2}\int_{-\infty}^{0}\!\! dy\int d^2b'
{\mathcal N}(b',y)\sin^2\!\Big(\frac{y\Phi(b')}{2}\Big) \label{intrep}
\end{eqnarray}
where the phase is
\begin{equation}
\Phi(b')=Gs\Big(\frac{2b'\cdot b}{b^2}+\log\frac{(b-b')^2}{b^2}\Big).
\end{equation}
The first term on the right hand side of eq. (\ref{intrep}) is from the tree-level high-energy eikonal phase $\partial\delta^{(0)}(b)/\partial b^{\mu} =-2Gs\,b^\mu/b^2$, 
\begin{equation}
-\frac{1}{m_1^2}\left(\partial\delta^{(0)}(b)/\partial b^{\mu}\right)^2 ~=~ -4\alpha_1^2 
\end{equation}
and the integral is what contains the $G$-resummation. To get this integral into a more convenient form we rescale $b'^{\mu}=|b| x^{\mu}$ and write
$\Phi=Gs\,\phi(x)$ with 
\begin{equation}
\phi(x)=2\hat{b}\cdot x+\log|\hat{b}-x|^2
\end{equation}
and $\hat{b}^{\mu} = b^{\mu}/|b|$. We likewise rescale rapidity to
$u=-m_1y|b'|=-m_1y|b||x|$ so that the argument of the $K_2$-function becomes $u$. and the argument of the $\sin$-function becomes
\begin{equation}
\frac{y\Phi(x)}{2}=-\frac{u}{2}\frac{Gs}{m_1b}\frac{\phi(x)}{|x|} = -\frac{u}{2} \alpha_1\frac{\phi(x)}{|x|}
\end{equation}
This finally gives
\begin{equation}
\frac{p_2^+}{m_1^2}\Delta p_1^-
= \!\!-4\alpha_1^{2} -\frac{4\alpha_1}{\pi^2}\int d^2x\;|x|^{-3}
F(\alpha_1\phi(x)/|x|),
\end{equation}
with   
\begin{equation}
F(z)=\int_0^\infty\! du\; u^2 K_2(u)^2\sin^2(zu/2)
\end{equation}
and thus manifestly making this a function of $\alpha_1$. It is worth stressing that this integral representation of $\Delta p_1^{-}$ is finite: the integral is bounded by the $\sin^2(zu/2)$-factor, the $K_2(u)$ is exponentially damped at large value of $u$, and the 2-dimensional integral over $x$ diverges neither close to the origin, nor close to $\hat{b}$. This holds irrespective of the magnitude of $\alpha_1$.

We find it convenient to do the $u$-integral first. Expanding $\sin^2(zu/2)$ in its convergent Taylor series we can integrate $K_2(u)^2$ over the resulting moments to get

\begin{eqnarray}
F(z) &=& \sum_{k=1}^{\infty}\frac{(-1)^{k+1}}{2\,(2k)!}\,z^{2k}\int_0^\infty\!u^{2k+2}K_2(u)^2\,du \cr
&=& \sum_{k=1}^{\infty}\frac{(-1)^{k+1}}{2\,(2k)!}\,z^{2k} W_{2k+2}
\end{eqnarray}
where
\begin{equation}
W_{2n+2}
  =\frac{2^{2n}}{(2n+2)!}\;\Gamma\!\Big(n+\frac72\Big)\,\Gamma\!\Big(n+\frac32\Big)^{2}\,\Gamma\!\Big(n-\frac12\Big).
\label{Wanalytic}
\end{equation}
The first few coefficients read $W_4=\frac{315}{512}\pi^2, W_6=\frac{4725}{4096}\pi^2$.

We can thus write
\begin{equation}
\frac{p_2^+}{m_1^2}\Delta p_1^- \!= \!-4\alpha_1^2
-\frac{2}{\pi^2}\sum_{k\ge1}\frac{(-1)^{k+1}}{(2k)!}\,W_{2k+2}\,X_{2k}\,\alpha_1^{2k+1}
\label{WX}
\end{equation}
where
\begin{equation}
X_{2k} =  \int d^2x\,|x|^{-(2k+3)}\phi(x)^{2k}
\label{X2kdef}
\end{equation}
The final step is to do the 2-dimensional (transverse) integral over $x$. This integral looks simple, and it {\em is} simple except for an increasing number of delicate cancellations which occur as the order of the expansion grows. By use of dimensional regularization away from 2D or, even simpler, by just replacing the logarithm in the phase $\phi(x)$ with the limit of a power, the master integral is the massless bubble integral
\begin{eqnarray}
B(\alpha,\beta)&=&\int d^2x\,(x^2)^{-\alpha}\big(|\hat{b}-x|^2\big)^{-\beta} \cr
  &=& \!\!\pi\,\frac{\Gamma(1-\alpha)\,\Gamma(1-\beta)\,\Gamma(\alpha+\beta-1)}
             {\Gamma(\alpha)\,\Gamma(\beta)\,\Gamma(2-\alpha-\beta)}    
\end{eqnarray}
a generalized Euler Beta-function. By the trick just mentioned, the coefficients we seek are thus composed of a finite sum of derivatives with respect to $\beta$ followed by evaluations at integer $\beta$-values (to generate the powers of logarithms). Finding any coefficient in the expansion is thus reduced to straightforward $\Gamma$-function algebra, and it can be done analytically to any desired order. To illustrate, 
let us consider the first coefficient $X_{2n}$ and here we introduce the notation $l_4 = \log(4)$. The relevant integrals all have single
logarithms and hence follow by just one differentiation of the Beta-function :
\begin{eqnarray}
\int\frac{\log w}{|x|^{5}}\!\!\!&=&\!\!\!\frac{4\pi}{9},~~
  \int\frac{\log w}{|x|^{3}}=4\pi \cr
  \int\frac{w\log w}{|x|^{5}}\!\!\!&=&\!\!\!\frac{16\pi}{9},~~
  \int\frac{(\log w)^2}{|x|^{5}}=\frac{16\pi}{9}\big(l_4-2\big), \nonumber
\end{eqnarray}
and they combine into
\begin{eqnarray}
X_2&=&2\Big(\frac{4\pi}{9}+4\pi-\frac{16\pi}{9}\Big)+\frac{16\pi}{9}(l_4-2) \cr
     &=&\frac{16\pi}{9}\,\big(l_4+1\big).
    \end{eqnarray}
which is the first non-trivial term of the expansion. Note that this coefficient already includes $l_4$ (times the overall $\pi$): a single transcendental quantity of weight $1 = 2n -1$; more on this below. As $n$ increases the derivatives distribute themselves by the Leibniz rule to give terms including polygamma functions such as $\psi^{(0)}(\frac12)=-\gamma_E-2\log2\ = -\gamma_E - l_4, \psi^{(2)}(1)=-2\zeta_3$, etc. Interestingly, all terms of kinds $\gamma_E, \pi^2 (= 6\zeta_2)$ cancel out. This may seem surprising (the $\gamma_E$'s cancel but $l_4$'s remain) and it is tied to the very specific values we need in the arguments of the Beta function. Specifically, $\alpha$ is always half-integer and $\psi^{(0)}({\rm half\!-\!integer}) = -\gamma_E - l_4 + \ldots$ (where the omitted terms are rationals) while $\psi^{(0)}({\rm integer}) = - l_4 + \ldots$, and their combinations are such as to cancel the $\gamma_E$'s. The same phenomenon makes even $\zeta$'s cancel. It is straightforward to prove that the $n$th term in the expansion is of maximum transcendentality $2n-1$ \cite{Moch:2001zr}. Indeed, the leading $\zeta$-term $\zeta^{2n-1}$ has a non-vanishing coefficient for all $n$ which we compute below.

 Evaluating the $X_{2n}$-terms and plugging them into eq. (\ref{WX}), we find complete agreement with the 11'th-order expansion of $\Delta p_1^-$ given in \cite{Saavedra:2026ttk} except for the coefficient 640 of $\ell_4\zeta_3^2$ at order $\alpha_1^9$ which should read 630. The analytical expression at order $\alpha_1^{13}$ is already too unwieldy to display here (it has 55 different combinations of powers of $\ell_4$ and odd $\zeta$-values). While we thus can push the calculation analytically to any desired order, it is clear that sheer complexity of the coefficients does not make it illuminating. Instead, we will now focus on the structure of this infinite-order expansion.

Given the physics input it is perhaps not surprising that there are recursive structures in the $\alpha_1$-expansion beyond the single even term proportional to $\alpha_1^2$. It is therefore convenient to focus on the integral $\mathcal I(\alpha_1) = p_2^+\Delta p_1^-/m_1^2 +4\alpha_1^2$ To illustrate, we can organize the coefficient of $\alpha^{2k+1}$ in powers of $\ell_4$ as follows:
\begin{equation}
\mathcal I(\alpha_1)=\pi\sum_{k=1}^{\infty}
\Big[\,A_k l_4^{2k-1}+B_k l_4^{2k-2}+\cdots\Big]\,\alpha_1^{2k+1} 
\end{equation}
This is not extracting the numerically largest contributions but it gives interesting analytical insight. With a complete abuse of language we can call this a "leading $\log(4)$ approximation", followed by next-to-leading, etc. For the two coefficients displayed above we find
\begin{eqnarray}
A_k &=& -\frac{2(-1)^{k+1}}{(2k)!}\,w_k\,c_k \cr
B_k &=& A_k\,(2k-1)\big(1-2\,O_{k-1}\big),
\end{eqnarray}
where $w_k = W_{2k+2}/\pi^{2}$ and
\begin{eqnarray}
c_k &=&\frac{k\,4^{\,k+1}}{(2k+1)^{2}} \cr
O_{k-1} &=& \sum_{j=1}^{k-1}\frac{1}{2j-1},\quad{\rm and}~~ O_0=0.
\end{eqnarray}
We could alternatively aim for recursive structures that ignore all logarithms and instead order the expansion according to the highest $\zeta$-values,
\begin{equation}
\mathcal I(\alpha_1) = \pi\sum_{k\ge2} [C_k\zeta_{2k-1}\,\alpha_1^{2k+1} + \ldots] .
\end{equation}
Here we find the simple relation
\begin{equation}
C_k = 8(-1)^{k+1}\,w_k\,\frac{2^{2k}-4}{(2k-1)(2k+1)^{2}}
\end{equation}
In this "highest $\zeta$-approximation" we thus have
\begin{eqnarray}
&&\frac{\mathcal I(\alpha_1)}{\pi} = 
+\frac{189}{128}\zeta_{3}\,\alpha_1^{5}
 - \frac{66825}{4096}\zeta_{5}\,\alpha_1^{7} \cr
 &&+ \frac{14189175}{32768}\zeta_{7}\,\alpha_1^{9}
 -\;\frac{11512580625}{524288}\zeta_{9}\,\alpha_1^{11} \cr
 &&+ \frac{7687645898625}{4194304}\zeta_{11}\,\alpha_1^{13} \cr
 &&-\;\frac{61314431790236625}{268435456}\zeta_{13}\,\alpha_1^{15}
+ \cdots 
\end{eqnarray}
Truncating at 11th order this matches with the corresponding terms given in \cite{Saavedra:2026ttk}. There are recursive relations also for $\zeta$-terms of lower weight but they are more complicated.

\section{Convergence properties of the all-order expansion}

An important issue to settle is whether the series in $\alpha_1$ converges in a specific range or has zero radius of convergence. As a prologue to this discussion we can consider an alternative evaluation of the $u$-integral that does not first expand in $\alpha_1$ but simply uses the identification with a generalized hypergeometric function:
\begin{eqnarray}
    F(z)&=&\!\int_0^\infty\! u^2K_2(u)^2\sin^2\frac{zu}{2}\,du \cr
&=& \!\!\!\frac{15\pi^2}{64}\Big[\,{}_3F_2\big(\frac72,\frac32,-\frac12;\frac12,2;-\frac{z^2}{4}\big)-1\Big].
\end{eqnarray}
The remaining integral over $x$ is now hard but we notice that the ${}_3F_2$-function has a branch point at unity, $i.e.$ when $z = \pm 2i$ and the point $\alpha_1 \sim 1$ is thus singled out. But there are no singularities on the real axis, corresponding to real $\alpha_1$. 

While the simple analytical expressions for the leading $log(4)$'s (whose individual infinite sums are finite numbers) might hint at convergence it is not difficult to see that the full sum of the $\alpha_1$-expansion has zero radius of convergence. To this end, let us write
\begin{eqnarray}
\mathcal I(\alpha_1) &=& \sum_{k=1}^{\infty}c_k\,\alpha_1^{2k+1} \cr
  c_k &=& -\frac{2}{\pi^2}\,\frac{(-1)^{k+1}}{(2k)!}\,W_{2k+2}\,X_{2k},
\end{eqnarray}
As we will explain in detail below, a large-order estimate gives
\begin{equation}
  c_k   \sim\ (-1)^k\frac{(2k+1)!}{4\,e^2\,4^{k}} ~~,~~ k \to \infty
  \label{ckasymptotic}
\end{equation}
which shows that (1) the series has zero radius of convergence and (2) it is expected to be Borel summable. The Borel transform divides out the factorial $(2k+1)!$ to form
\begin{equation}
B(t) = \sum_k\frac{c_k}{(2k+1)!}t^{2k+1}
\label{Bdef}    
\end{equation}
with leading-order estimate $B(t)\sim t/(1+t^2/4)$, and $t=\pm 2i$ are actually end-points of branch cuts. We thus expect the Borel sum to be convergent inside a radius of length two.

Before proceeding to any detailed calculation, It is of interest to disentangle where the asymptotic behavior of factorial growth in the coefficients arises. Individual terms in the coefficients $X_{2n}$ do not display factorial growth but the proliferation of terms (the $l_4$'s and odd $\zeta$'s and their products) grows factorially. This factorial growth is compensated by the $1/(2k)!$ behavior of the expanded $\sin^2(zu/2)$ function, and in this counting the origin is the factorial growth from the integral over moments of $K_2(u)^2$ inside the $u$-integral. This makes it clear that it is the (convergent) Taylor expansion of $\sin^2(zu/2)$ in the integrand followed by a term-by-term integration that causes the series to diverge. What will make the series Borel summable is thus the oscillating sign $(-1)^k$. 

Let us now derive these asymptotic estimates in all detail. The two ingredients are the (squared) Bessel-$K$ moments
\begin{equation}
  W_n = \int_0^{\infty} t^{\,n}\,K_2(t)^2\,dt,
\label{eq:Wdef}
\end{equation}
and the two-dimensional integral $X_{2k}$ of eq. (\ref{X2kdef}).
Both of these integrals are finite: $X_{2k}$ converges at the origin (where $\varphi \sim |x|^2$), at
infinity (where the integrand falls off as $|x|^{-3}$), and at the point $x=\hat b$ where we encounter the integrable logarithmic singularity which is most easily dealt with
analytically by means of taking the limit of a power, as explaned above. Finally, it is known that $W_n$ converges
for all $n=2k+2$ with $k\ge1$.

Applying Stirling's approximation to the exact analytical expression (\ref{Wanalytic}) gives
\begin{equation}
W_{2k+2} \sim \frac{\pi}{2}\,\frac{(2k+1)!}{2^{2k+2}}
\end{equation}
which forms the first part of our asymptotic estimate. Next, we need a corresponding asymptotic estimate of the two-dimensional integral $X_{2k}$. Clearly, the asymptotic behavior
of this factor is dominated by the region around $x=\hat b$, the only place where $\varphi$ is unbounded. Introducing polar coordinates $(r,\psi)$ such that
$|\hat b-x|=r$, $2\hat b\!\cdot\! x = 2+2r\cos\psi$, $|x|^2=1+2r\cos\psi+r^2$ and
$d^2x=r\,dr\,d\psi$, an exact rewrite gives
\begin{equation}
  X_{2k} = \!\int_0^{2\pi}\!\! d\psi \int_0^{\infty}\!\!\! dr
  r(1+2r\cos\psi+r^2)^{-(2k+3)/2}\,\varphi^{\,2k}
\label{Xpolar}
\end{equation}
with $\varphi = 2 + 2r\cos\psi + 2\log r$. As expected,
the radial integrand is sharply peaked, and this peak moves towards the endpoint as $k$
grows. Extremizing $\log r + 2k\log|2+2\log r|$ (the $\psi$--independent part) gives
$1+\log r_0 = -2k$, $i.e., r_0 = e^{-(2k+1)}$, 
the saddle point we will use to estimate the integral. At this point, which moves towards $r=0$ as $k$ grows, the other regions of the integral remain exponentially
negligible. We can therefore safely replace $2\hat b \cdot x$ with the constant 2 (its endpoint value), and
$(1+2r\cos\psi+r^2)^{-(2k+3)/2}$ with unity. In this limit the angular integral trivializes, leaving just a flat factor of $2\pi$, but care must be taken to not impose these
simplifications in the integrand outside the region of validity. We therefore cut off the radial integration at some finite value, here for convenience chosen to be unity.
The radial integral will then becomes elementary after
a simple change of variables,
\begin{equation}
  \int_0^1 dr r (2+2\log r)^{2k}
  \!= \! \frac{1}{2}\,e^{-2}\!\int_{-2}^{\infty}\! t^{2k}e^{-t} dt ~.
\end{equation}
The $t$-integral has two parts: the interval $[-2,0]$ and the interval $[0,\infty]$. The latter dominates exponentially over the former, and we can
asymptotically replace
\begin{equation}
\frac{1}{2}\,e^{-2}\!\int_{-2}^{\infty}\! t^{2k}e^{-t} dt \sim \frac{1}{2}\,e^{-2}\!\int_{0}^{\infty}\! t^{2k}e^{-t} dt
\end{equation}
which evaluates to $(1/2)e^{-2}\,(2k)!$, and thus
yields the asymptotic estimate $X_{2k} \sim \pi\,e^{-2}\,(2k)!$. Combined with the asymptotic estimate of $W_{2k}$ this gives the asymptotic limit of coefficients $c_k$ as given in eq. (\ref{ckasymptotic}) above. 

We have checked numerically
that the asymptotic estimate (\ref{ckasymptotic}) indeed seems to be approached as $k$ grows, although the convergence towards this asymptotic behavior is slow. This is shown
in Table 1, where we compare numerical values $c_k^{num}$ with the asymptotic estimate of eq. (\ref{ckasymptotic}). Even at order $\alpha_1^{25}$ the asymptotic estimate
still differs by approximately 34\%. 

\begin{table}[t]
\centering
\begin{tabular}{r l r r }
\toprule
$k$ & order & $c_k^{num}$ & $c_k^{num}/c_k$  \\
\midrule
$1$  & $\alpha_1^{3}$  & $-8.19959$                 & $161.566$  \\
$2$  & $\alpha_1^{5}$  & $+4.11893$                 & $16.2320$  \\
$3$  & $\alpha_1^{7}$  & $-10.5143$                 & $3.94619$  \\
$4$  & $\alpha_1^{9}$  & $+109.196$                 & $2.27683$  \\
$5$  & $\alpha_1^{11}$ & $-2.51002\times10^{3}$     & $1.90314$  \\
$6$  & $\alpha_1^{13}$ & $+8.86511\times10^{4}$     & $1.72351$ \\
$7$  & $\alpha_1^{15}$ & $-4.33748\times10^{6}$     & $1.60623$  \\
$8$  & $\alpha_1^{17}$ & $+2.79495\times10^{8}$     & $1.52207$  \\
$9$  & $\alpha_1^{19}$ & $-2.28989\times10^{10}$    & $1.45851$  \\
$10$ & $\alpha_1^{21}$ & $+2.32235\times10^{12}$    & $1.40875$  \\
$11$ & $\alpha_1^{23}$ & $-2.85429\times10^{14}$    & $1.36871$  \\
$12$ & $\alpha_1^{25}$ & $+4.17851\times10^{16}$    & $1.33581$ \\
\bottomrule
\end{tabular}
\caption{The relatively slow convergence towards the asymptotic behavior of coefficients $c_k$.}
\end{table}

The slow approach is easily understood as being due to the limited accuracy of repeated use of Stirling's approximation
in the $\Gamma$-functions of the exact analytical expression for $W_{2k+2}$. A more accurate estimate for $W_{2k+2}$ can be obtained by working at the integrand level
instead. Using the asymptotic expansion of the Bessel $K_2$-function and keeping the first sub-leading term corresponding to $K_2(u)^2=\frac{\pi}{2u}e^{-2u}(1+\frac{15}{4u}+\ldots)$ gives
\begin{equation}
W_{2k+2} \sim \frac{\pi}{2}\,\frac{(2k+1)!}{2^{2k+2}}\left(1 + \frac{15}{2(2k+1)}\right)
\end{equation}
after discarding the higher-order corrections. In Table 2 we show how the improved estimate for a selection of powers of $\alpha_1$ based on including the sub-leading correction, 
\begin{equation}
  c_k^{NL}   \sim\ (-1)^k\frac{(2k+1)!}{4\,e^2\,4^{k}}\left(1 + \frac{15}{2(2k+1)}\right)
\end{equation}
leads to a deviation of less than 3\% at order $\alpha_1^{25}$.

\begin{table}[t]
\centering
\begin{tabular}{r r r r}
\toprule
$k$ & order & $c_k^{num}/c_k$ & $c_k^{num}/c_k^{NL}$ \\
\midrule
$3$  & $\alpha_1^7$ & $3.94619$ & $1.90506$  \\
$5$  & $\alpha_1^{11}$ & $1.90314$ & $1.13160$ \\
$8$  & $\alpha_1^{17}$ & $1.52207$ & $1.05613$ \\
$10$ & $\alpha_1^{21}$ &  $1.40875$ & $1.03802$ \\
$12$ & $\alpha_1^{25}$ & $1.33581$ & $1.02755$ \\
\bottomrule
\end{tabular}
\caption{Including the first sub-leading correction gives clear evidence of the validity of the inferred asymptotic behavior.}
\end{table}

Which function does the Borel summed expression converge to? In contrast to the usual situation in physics we already {\em know} that full function since we precisely obtained the series expansion from it through an improper combination of Taylor expansion of the integrand followed by integration. The Borel sum is thus the integral $\mathcal I(\alpha_1)$ but a priori only valid inside the Borel radius of 2. However, as we have stressed above, the integral $\mathcal I(\alpha_1)$ is well defined on the whole real positive $\alpha_1$-axis. This removes the condition on $\alpha_1$ to be less than unity. There is no contradiction in this since the integral $\mathcal I(\alpha_1)$ is well-defined for all finite real $\alpha_1$ to begin with. It should be stressed that this does not imply that $\mathcal I(\alpha_1)$ is guaranteed to be physically meaningful for all $\alpha_1$ since it was derived from weak-coupling perturbation theory.

It is of interest to see how well these expectations are borne out in detail by comparing numerically the Borel-summed series (\ref{Bdef}) with the original expression $\mathcal I(\alpha_1)$. To this end, let us write $B(t) = t g(v)$ with $v \equiv t^2$ so that the Borel sum is given by
\begin{equation}
g(v)=\sum_{k=1}^{\infty}\beta_k\,v^{k}  \beta_k=\frac{c_k}{(2k+1)!}
\label{gsum}
\end{equation}
where $c_k$ are the {\em exact} coefficients. As we have derived, the asymptotic estimate of the $\beta_k$-coefficients is $\beta_k \sim (-1)^k/(4e^2\,4^k)$, and $g(v)$ thus 
has radius of convergence 4, with its nearest singularity at $v=-4$ (the image of $t=\pm 2i$). However, the analytically continued function allows us, in a standard manner, to
define the Laplace transform
\begin{equation}
 {\mathcal J}(\alpha_1)=\int_0^{\infty} ds e^{-s} \alpha_1 s g\big((\alpha_1 s)^2\big) ~.
\label{Jdef}
\end{equation}
Clearly, if we simply insert the series expansion of $g(v)$ we only recover the asymptotic series which was the starting point. Instead, we map the cut $v$-plane to a disk by
$v = 16w/(1-w)^2$, corresponding to
\begin{equation}
w(v)=\frac{v+8-4\sqrt{v+4}}{v} ~.
\label{wofv}
\end{equation}
The cut on the negative $v$-axis from -4 to minus infinity is mapped to the unit circle and the real $v$-axis is mapped to the unit interval in $w$.
Substituting the map of $v$ to $w$ into (\ref{gsum}) and expanding in $w$ we obtain $g=\sum_{m}a_m w^m$ with coefficients 
\begin{equation}
 a_m=\sum_{k=1}^{m}\beta_k 16^{k} \binom{k+m-1}{m-k} 
\end{equation}
If we had all coefficients $c_k$ available this expansion substituted into eq. (\ref{Jdef}) would give us the full Borel-summed function. In practice, we have only exact analytical
expressions for a finite number of coefficients but it is still illuminating to see the result of keeping only such a finite number of coefficients and evaluate the integral
(\ref{Jdef}) numerically. In Figure 1 we show the result of such numerical integrations.  

\begin{figure}[h]
\centering
\includegraphics[width=\columnwidth]{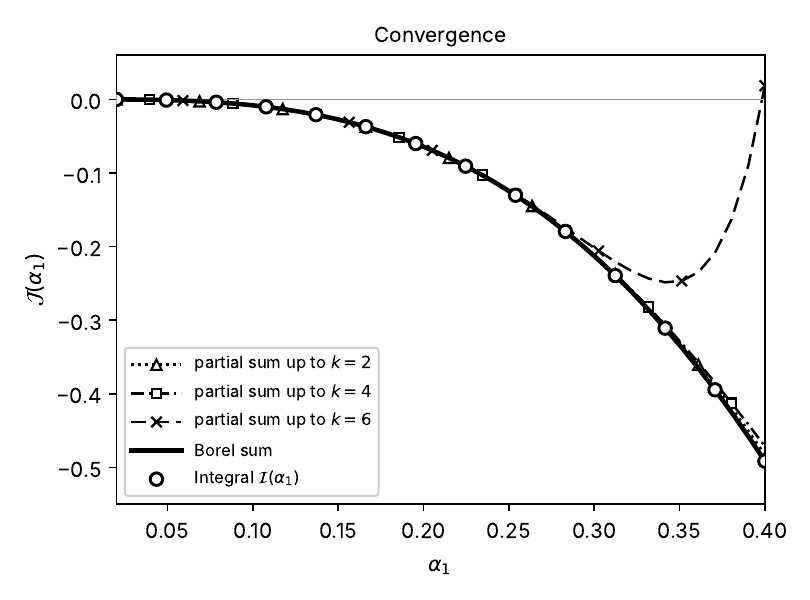}
\caption{Plot showing first how the simple summations of first few terms work well for small values of $\alpha_1$ but eventually diverge because the sum is asymptotic only.
Note that the ordering is such that including more terms make the sums diverge for smaller values of $\alpha_1$. In contrast, the Borel-summed expression based on the
analytic continuation to the full real positive axis by means of the conformal map is indistinguishable from the full integral even when only based on the first twelve terms of
the expansion.}
\end{figure}

\section{Conclusion}
We have explored the convergence properties of the recently proposed all-order resummation of high-energy gravitational scattering in the rescaled gravitational coupling $\alpha_1$ \cite{Saavedra:2026ttk}. The expansion turns out to be asymptotic only, but we have provided evidence that the series is Borel summable and with a Borel-radius of two. The Borel sum will thus agree with the integral representation, and interestingly this integral representation is finite for all values of $\alpha_1$ on the real positive axis. This is indicative of a change of regime at $\alpha_1 \sim 1$ (as expected on physical grounds) and it could suggest a phenomenon of resurgence in high-energy gravitational scattering in this specific high-energy regime, potentially paving the way towards strong-coupling calculations for $\alpha_1 \geq 1$ that might bypass the series expansion entirely.

The asymptotic series does have a number of interesting properties that suggest it has deeper roots than the detailed diagrammatic derivation would indicate. This is evidently not an uncontrolled approximation of unknown accuracy: it has passed a number of tests based on existing fixed-order post-Minkowskian calculations that expands in $G$ \cite{Saavedra:2026ttk}, and as we have pointed out here the analytic properties of the series expansion contains beautiful structures, of which only some of those are responsible for Borel summability. Many questions are thus raised by this new approach to a high-energy gravitational scattering regime which deserve further attention.






\end{document}